\documentclass[aps,amsfonts,amsmath,prd,preprint,nofootinbib]{revtex4}
\usepackage{epsf,mathrsfs}

\newcommand{\beq}{\begin{equation}}
\newcommand{\eeq}{\end{equation}}

\begin{document}

\title{Holographic multiverse}

\author{J. Garriga$^1$ and A. Vilenkin $^2$}
\address{
$^1$ Departament de F{\'\i}sica Fonamental i \\Institut de Ci{\`e}ncies del Cosmos, 
Universitat de Barcelona,\\
Mart{\'\i}\ i Franqu{\`e}s 1, 08028 Barcelona, Spain\\
$^2$ Institute of Cosmology, Department of Physics and Astronomy,\\
Tufts University, Medford, MA 02155, USA}

\begin{abstract}

We explore the idea that the dynamics of the inflationary
multiverse is encoded in its future boundary, where it is described by
a lower dimensional theory which is conformally invariant in the
UV. We propose that a measure for the multiverse, which is needed in
order to extract quantitative probabilistic predictions, can be derived 
in terms of the boundary theory by imposing a UV cutoff. In
the inflationary bulk, this is closely related (though not identical)
to the so-called scale factor cutoff measure.

\end{abstract}

\maketitle

\section{Introduction}

In a theory with many low-energy vacua, the dynamics of eternal
inflation generates a variety of environments where observers may
develop and experiments may take place.  One would then like to
calculate probability distributions for such environments, including,
for instance, probabilities for what we commonly call the cosmological
parameters of the standard Big Bang model, or the parameters of the
low energy standard model of particle physics (for recent reviews, see
e.g. \cite{inflation,landscape}).

Suppose the universe starts in an inflationary false vacuum with a
sufficiently low decay rate.  Localized bubbles of neighboring vacua
will occasionally nucleate and subsequently expand into the false
vacuum. Each bubble takes a small fraction of the volume in the
original inflating vacuum into a new phase, but since the parent
vacuum is inflating, the daughter bubbles never deplete it
completely. Some of the vacua within these daughter bubbles may also
inflate, decaying into other vacua, and so on. This leads to a never
ending cascade of bubbles within bubbles, where all possible phases in
the theory are eventually realized.

Here, and for most of the paper, we assume that the ``landscape" of
vacua is irreducible.  This means that we can access any given
inflationary vacuum from any other one through a finite sequence of
transitions. The transitions or ``decays" can happen from high energy
inflationary vacua to low energy inflationary vacua, but also in the
opposite direction.  If the landscape is not irreducible, then for
present purposes we may think of the different irreducible sectors as
different theories. In an irreducible landscape, the volume
distribution of the different vacua quickly approaches an attractor
solution as a function of time \cite{GSVW}.  The explicit form of the
volume distribution depends on which time coordinate we use in order
to slice the inflationary ``multiverse", but the important point is
that the attractor is independent of the initial
state.\footnote{Nonetheless, inflation is not eternal to the past
  \cite{Borde:2001nh}.  It must start somehow, and a description of
  the beginning of inflation may be needed for logical
  completeness. For instance, in the case where the landscape of the
  theory is reducible, this may help addressing the question of which
  irreducible sector are we more likely to be in.}

In the multiverse, all localized physical processes which are allowed
by the theory will happen an infinite number of times. This suggests
that probabilistic predictions should be based on a statistical
counting of occurrences of the different kinds of processes we may be
interested in \cite{Garriga:2007wz}. However, the results of this
counting depend very much on how we regulate the infinities, and
different regulators give vastly different results for the
probabilities \cite{LLM}. This is the so-called ``measure problem" of
inflationary cosmology.

There is by now a fairly extensive literature of proposals for the
measure, some of which have already been superseded or simply
abandoned because of inherent paradoxes or inconsistencies
\cite{measure}.  The fact remains, however, that in order to
confront theory with observations the measure is needed, and the
question must be addressed.  Progress has been made in recent years,
by way of identifying some generic problems which a measure should
avoid. Amongst those we may highlight the following:

{\em The youngness paradox.} This corresponds to the prediction that
we are exponentially more likely to have emerged much earlier in
cosmic history, in strong conflict with observations
\cite{Guth2000,Tegmark,youngness2}.  This problem afflicts the
so-called ``global proper-time cutoff measure", which is defined as
follows. First, we choose an arbitrary spacelike surface
$\Sigma$, which serves as the origin of proper time along a congruence
of geodesics orthogonal to it. In the ensemble of all events, we count
only those which happen before a fixed proper time.  The probabilities
are calculated in terms of this counting, in the limit when the proper
time cutoff is sent to infinity. The same "youngness" effect arises
if we use some other time variable in order to impose the global
cutoff. Nevertheless, there is a narrow class of choices of the
cutoff time variable where the ``paradox" turns into a quite benign
``youngness bias" \cite{DGSV}. The scale-factor cutoff measure, to be
discussed below, belongs to this class.

{\em The Boltzmann brain paradox.} This corresponds to the prediction
that we are most likely to be disembodied brains, created by large
quantum fluctuations in an otherwise empty phase, dreaming of CMB
multipoles and other exquisite data
\cite{Rees1,Albrecht,Page1,challenges,DKS02}.  This paradox afflicts a
broad class of measure proposals, and is due to the fact that
thermalized regions like ours (which are thought to arise as a
consequence of an earlier phase of slow roll inflation) are relatively
rare in the multiverse. Most of the space-time volume is occuppied by
empty quasi-de Sitter phases. Because of that, ``freak" observers,
created by large quantum fluctuations in vacuum, may easily outnumber
the ``ordinary" observers who live in thermalized regions. Of course,
this will depend on how we regulate the numbers of both types of
observers, and it is conceivable that some measures may avoid the
problem (provided that the landscape of vacua satisfies certain
properties). This issue is currently under active investigation
\cite{Bousso:2008hz,DeSimone:2008if}.

{\em The Q-catastrophe.} This afflicts proposals where probabilities
are rewarded according to the amount slow-roll inflationary expansion
which precedes thermalization, such as for instance, the
``pocket-based" measure introduced in \cite{GTV,pockets,GSVW}.  In
this case, observables which are correlated with the number of
e-foldings of slow-roll tend to suffer an exponential bias towards
large or small values \cite{catastrophe}. One such observable is the
amplitude of density perturbations $Q$ caused by quantum fluctuations
of the inflaton field. The exponential bias would push the likely
values of $Q$ towards the boundaries of what is anthropically allowed,
while the observed value happens to sit comfortably in the middle of
the anthropic range. There may be ways out of this ``catastrophe"
(such as the possibility that density perturbations are seeded by a
curvaton rather than the inflaton), but correlations of the parameters
with the duration of slow roll inflation remains a potential
nuissance.

These ``paradoxes" and ``catastrophes" can be thought of as
phenomenological constraints, useful to narrow down the possible
definitions of a measure. Additionally, we may take the point of view
that {\em initial conditions} at the beginning of inflation {\em
should be irrelevant} for the purposes of making predictions
\footnote{See, however, the caveat mentioned in the previous
footnote.}.  The justification is that the dominant part of the
spacetime volume in the eternally inflating ``multiverse" is in the
asymptotic future, where the volume distribution of different vacua
(and physical processes therein) is well described by an attractor
solution which is insensitive to initial conditions.

It has recently been emphasized
\cite{DGSV,Bousso:2008hz,DeSimone:2008if} that the scale factor
cutoff measure (which we shall discuss in Section IV) is free from
the youngness paradox, does not suffer from the Q-catastrophe, and is
independent of initial conditions.  It also provides a good fit to the
data when it is used in order to predict the likely values of the
cosmological constant \cite{DGSV} and, with some relatively mild
assumptions about the landscape, it avoids the Boltzmann brain paradox
\cite{Bousso:2008hz,DeSimone:2008if}.  It thus appears to be a
promising candidate for the measure of the multiverse.

The purpose of this paper is to formulate a measure for the
multiverse which may be connected to the underlying dynamical
theory. Our discussion relies on the existence of the late time
attractor for the volume distribution of vacua, and it is inspired by
the holographic ideas which have been developed in the context of
string theory. The idea is to formulate the calculation of
probabilities directly at the ``future boundary" of space-time, the
place where everything has been said and done. Data at the future
boundary contains information on everything that has happened in the
multiverse, and this makes it a natural locus to do our counting.

The plan of the paper is the following. In Section II we review the
causal structure of the inflationary multiverse, and formalize our
definition of the future boundary. In Section III we propose that the
dynamics of the multiverse can be mapped into a lower dimensional
Euclidean field theory, which lives at the future boundary.  Under
this mapping, the late time (or ``infrared") self-similar behaviour of
the attractor solution in the bulk, corresponds to scale invariance 
in the ``ultraviolet" of the boundary theory. Conversely, the
infrared properties of the boundary correspond to the initial stages
of inflation, which depend on initial conditions.  The measure is
discussed in Section IV.  By analogy with field theory, we propose to
regulate infinities at the boundary by imposing a UV cutoff.
Probabilities for different types of events are then defined as the
ratios of occurrences in the limit when the cutoff is removed.  As we
shall see, this procedure is closely related to the scale-factor
cutoff measure. Our conclusions are summarized and discussed in
Section V.

\section{The future boundary}

The future causal boundary, which we shall denote as $c^+$, can be
defined as the set of endpoints of inextendible time-like curves. More
precisely, ``points" or elements of $c^+$ are defined as the
chronological pasts of inextendible time-like curves. Two curves with the same
past will therefore define the same ``endpoint" at $c^+$\cite{HE}. The
future boundary contains points of various different types.

In the inflationary multiverse, most time-like curves will eventually
exit the inflating region and fall into one of the non-inflating
vacua, which we shall refer to as terminal vacua. Terminal vacua may
have vanishing or negative vacuum energy, and will be denoted as
Minkowski and anti-de Sitter (AdS) vacua respectively.  Bubbles of AdS
vacua develop a spacelike singularity or ``big crunch" in their
interior, and timelike curves hit this singularity in a finite proper
time. The corresponding endpoints will be said to belong to the
singular part of $c^+$. Time-like curves which enter a metastable
Minkowski vacuum will also fall, eventually, into one of the AdS
vacua, adding to the singular boundary. However, supersymmetric
Minkowski vacua are completely stable \cite{Deser}, and time-like
curves entering them will span the future null and time-like
infinities of the Minkowski conformal boundary, usually referred to as
$\mathscr{I}^+$ and $i^+$ respectively. Fig. 1 represents the causal
structure of an eternally inflating universe with terminal bubbles of
different types. The conformal future boundary of Minkowski bubles has
the shape of a ``hat" \cite{hat2}, whereas the spacelike singularity
at the future boundary of AdS bubbles is represented by a broken line.

\begin{figure}[ht]

\begin{center}\leavevmode  
\epsfxsize=16.5 cm
\epsfbox{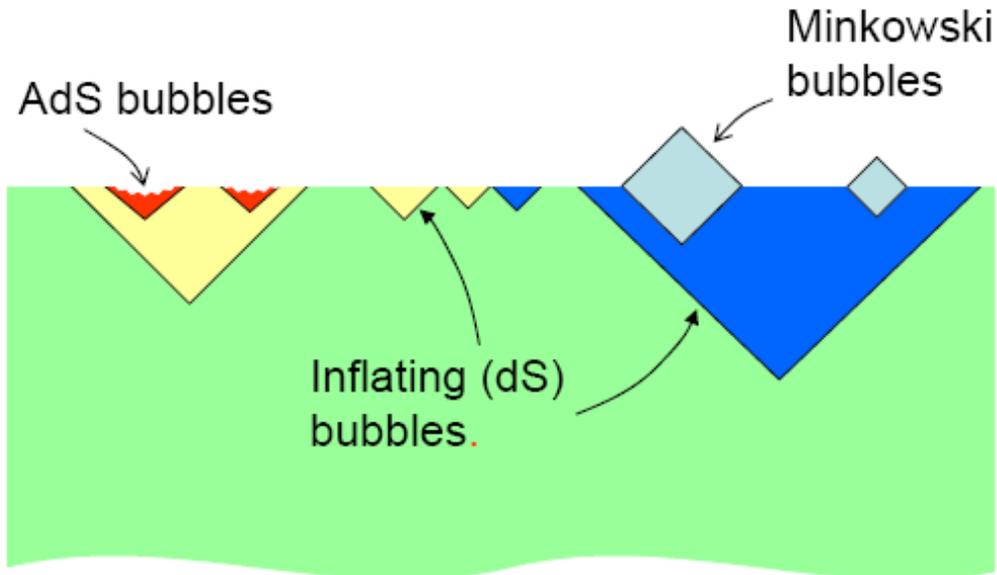}
\end{center}

\caption{Causal diagram of the inflationary multiverse. The vertical direction 
represents time, and the horizontal direction is space. Bubbles of different
types nucleate and start expanding close to the speed of light. Bubbles with positive 
vacuum energy (dS bubbles) inflate eternally. Inflation stops in 
bubbles with vanishing or negative vacuum energy (Minkowsi and AdS bubbles, respectively).}
\label{2j}
\end{figure}

On the other hand, because inflation is eternal, some time-like curves
will remain forever in inflating regions of space. The corresponding
endpoints will be called eternal points, and they can also be
classified in different types. For instance, there are time-like
geodesics which, after a finite sequence of transitions, remain
forever in a given inflating vacuum. If we assign a ``color" to each
inflating vacuum, the corresponding endpoints would have a definite
color. However, eternal time-like curves are more likely to jump back
and forth between different vacua on their way to infinity, without
ever settling into any of them. In this case, the corresponding
endpoints would have, so to speak, a mixed color. Clearly, there are
many different hues the eternal points may have.

The future boundary can be endowed with a topology, whose open sets
are defined in terms of the future of points in the manifold. We say
that a point $p\in c^+$ belongs to the open set $U_{int}(q)$, if the
time-like curves whose endpoint define $p$ have some intersection
with the chronological future of point $q$ in the manifold ${\cal
M}$. Intuitively, the intersection of the future lightcone of point
$q$ with $c_+$ draws the boundary of an open set $U_{int}(q)$ in
$c^+$.  Likewise, we say that $p$ belongs to the open set $U_{ext}(q)$
if its defining time-like curves do not intersect the {\em causal}
future of point $q$. By definition, arbitrary unions and finite
intersections of the $U_{int}(q)$ and the $U_{ext}(q')$ for all $q,
q'\in {\cal M}$, are also open sets. It is unclear to us whether one
can make $c^+$ into a differentiable manifold by using this topology.
For this we would need an invertible map from the open sets
$U_{int}(q)$ and $U_{ext}(q)$ onto ``coordinate" open sets of $R^3$,
in such a way that the changes of coordinates are differentiable where
open sets overlap. While this remains an interesting possibility, we
shall not pursue it here.  The reason is that we need not work with
the full set $c^+$, but only with the set of eternal points (and its
boundary), as we shall now describe.

\subsection{The fiducial spacelike hypersurface $\Sigma_3$}

Consider a space-like hypersurface $\Sigma_3$ embedded in the
multiverse and a congruence of time-like geodesics $\gamma({\bf x})$
orthogonal to it (see Fig. 2).  Here $\bf x$ are coordinates on
$\Sigma_3$.  It is not necessary that $\Sigma_3$ be a Cauchy surface
for the whole multiverse, but we do require that at least one of the
geodesics in the congruence be eternal. This guarantees that the
attractor solution is reached at late times in the spacetime region
${\cal S}$ spanned by the congruence. Because of that, the portion of
$c^+$ attached to this region will be a fair sample of the whole of
$c^+$.

\begin{figure}[ht]

\begin{center}\leavevmode  
\epsfxsize=16.5 cm
\epsfbox{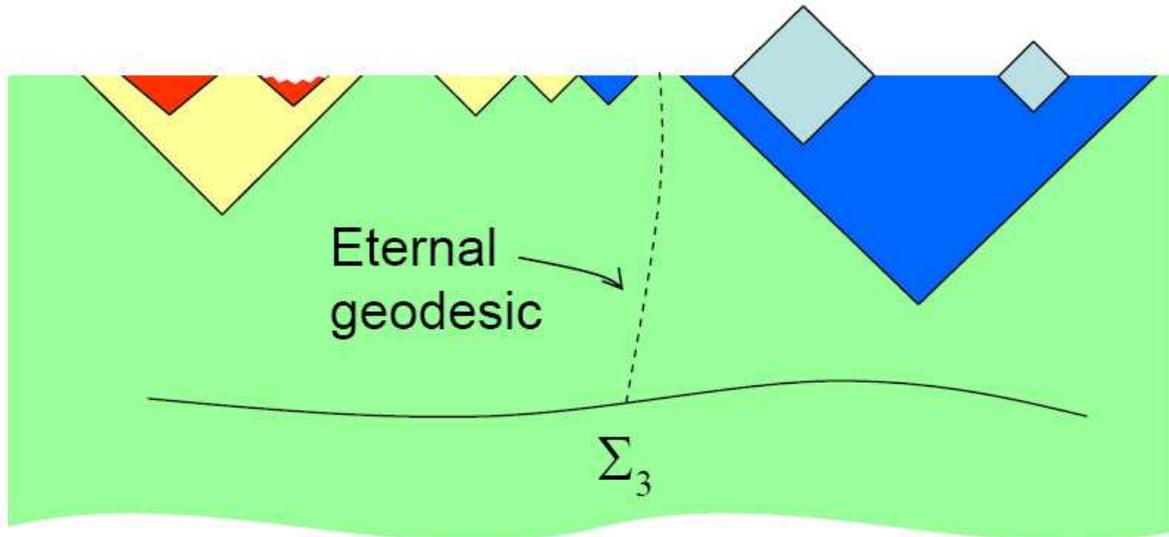}
\end{center}

\caption{We assign co-moving coordinates to the points in the eternal
set $E$, by introducing a fiducial hypersurface $\Sigma_3$ and a
congruence of geodesics orthogonal to it. The metric $g_{ij}({\bf x})$
on $\Sigma_3$ can be used in order to assign co-moving distances
amongst the points in $E$.}
\label{4j}
\end{figure}

The congruence defines co-moving coordinates in ${\cal S}$, and the
metric $g_{ij}({\bf x})$ induced on $\Sigma_3$ defines co-moving
distances between different geodesics. Bubbles nucleating to
the future of $\Sigma_3$ can be projected backwards along the
congruence. The ``image" of a bubble consists of all those points
${\bf x}$ in $\Sigma_3$ such that $\gamma({\bf x})$ intersects the
given bubble.  The congruence also allows us to define ``scale factor
time" $a$. At any spacetime point $x$ in ${\cal S}$ this is given by
\begin{equation}
\ln a(x)= \int_{{\bf x}(x)}^x H d\tau.\label{scalefactor}
\end{equation}
Here $\tau$ is proper time and the integral is taken along the
geodesic in the congruence that connects $x$ to some point ${\bf
x}(x)$ on $\Sigma_3$, while $H$ is one third of the expansion
$\nabla_{\mu} u^\mu$ of the congruence. Here $u^\mu=dx^\mu/d\tau$ is
the tangent vector. Note that with this definition, 
\begin{equation} 
a(x\in \Sigma_3)=1.  
\end{equation}
The co-moving volume of the image of a bubble of vacuum $i$
nucleating in inflating vacuum $j$ at scale factor time $a$ is given
by \cite{GSVW}
\begin{equation}
V_i (a) = {4\pi\over 3} H_j^{-3} a^{-3}, \label{cv}
\end{equation}
where $H_j=(\Lambda_j/3)^{-1/2}$ is determined by the effective vacuum energy $\Lambda_j$.
Hence, bubbles nucleating later in time will have smaller images on
$\Sigma_3$. 

Here, and for the rest of the paper, we disregard geodesic
crossing.  Of course, due to gravitational instability, structure will
form in some of the pocket universes and some geodesics will
eventually cross each other. To avoid this effect, we shall define
co-moving distances as is usually done in standard cosmology, by
considering geodesics on a metric which is smoothed out on sufficiently 
large scales. In this case, the congruence will always be diverging,
except in the collapsing AdS regions.  As we shall see, we will not
need to be specific about the details of the congruence inside of
terminal bubbles.

\begin{figure}[ht]

\begin{center}\leavevmode  
\epsfxsize=16.5 cm
\epsfbox{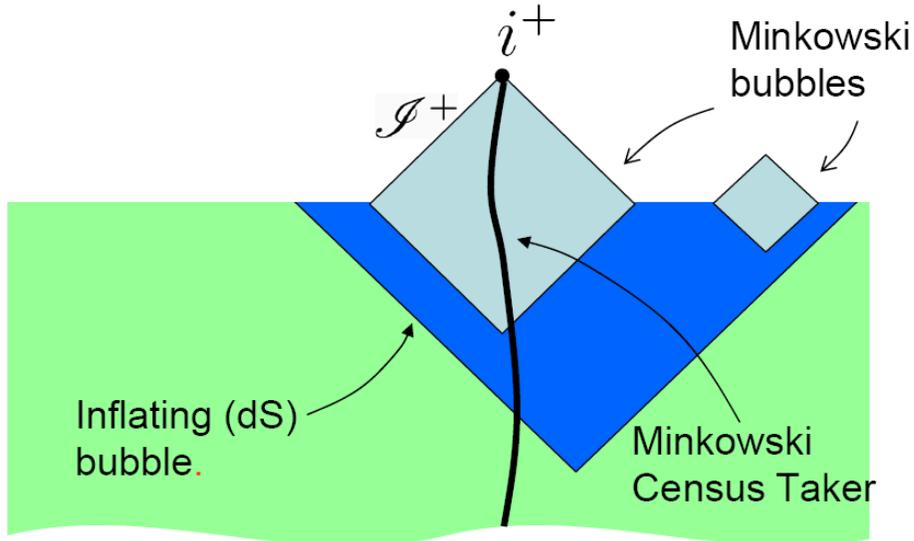}
\end{center}

\caption{The future boundary of a Minkowski bubble is a "hat", consisting of
the union of future null infinity $\mathscr{I}^+$ and time-like infinity $i^+$.
The worldline of a "census taker" ending at $i^+$ is also represented.}
\label{3j}
\end{figure}

Each geodesic $\gamma({\bf x})$ defines an ``endpoint" $p({\bf x})$ in
$c^+$, and so the congruence maps the fiducial hypersurface $\Sigma_3$
onto the future boundary. However, this map is not one to one. Indeed,
the image of a stable Minkowski bubble occupies a finite co-moving
volume on $\Sigma_3$, given by Eq. (\ref{cv}), while the
corresponding geodesics all end up at the same single point in $c^+$,
namely, the point $i^+$ of the conformal boundary of Minkowski.
Conversely, the future null infinity $\mathscr{I}^+$ of a bubble of
the stable Minkowski vacuum is not reached by any of the time-like
geodesics $\gamma({\bf x})$, and so it has no anti-image on
$\Sigma_3$. Fortunately, this will not necessarily be a problem.

The reason is that the map fails to be one to one only in the regions
corresponding to the ``hats" of stable Minkowski bubbles. However, it
has been argued in \cite{hat2,hat1} that the bulk region in the interior
of these bubbles is holographically described by a field theory which
lives on the 2 dimensional surface $\Sigma_2$ which lies at the
boundary of $\mathscr{I}^+$, where the hat meets the ``horizontal"
part of $c^+$ (see Fig. 3). In this sense, we need not worry about
points on $c_+$ which are inside of $\Sigma_2$.  This means that on
the fiducial hypersurface $\Sigma_3$ we can excise the images of
stable Minkowski bubbles: these will be accounted for by degrees of
freedom which live at the boundary of the excised holes. As a matter
of fact, we will also argue that the images of AdS bubbles,
corresponding to big crunch singular points, should be excised in a
similar way. In the case of AdS bubbles, the surface $\Sigma_2$ is the
boundary of the corresponding set of ``big crunch" singular points on
$c_+$.

\section{Holography}

In this Section we ellaborate on the idea that the dynamics of the
multiverse is dual to a boundary field theory. This is of course
inspired by the holographic AdS/CFT correspondence, and the idea of
applying it to inflationary cosmology is not new. The version we
advocate here builds up from two earlier proposals.

In \cite{strominger} it was proposed that the dynamics of de Sitter
space is dual to a Euclidean CFT which lives at the conformal future
boundary. That construction, however, did not allow for the
possibility that a given de Sitter phase decays into neighboring
vacua. Such decays drastically modify the structure of the future boundary, 
making it very different from the conformal boundary of de Sitter.

A somewhat related proposal was developed in
Refs. \cite{hat1,hat2,hat3}. There, it is argued that the bulk
dynamics of a pocket universe corresponding to a stable Minkowski
vacuum is dual to a 2-dimensional Euclidean field theory, which lives
at the boundary $\Sigma_2$ of the Minkowski ``hat" representing the
future null infinity of that pocket. For a single bubble, this
boundary has the topology of a 2-sphere. The field theory on
$\Sigma_2$ encodes the information corresponding to any observation
that can be made by a hypothetical ``census taker" who lives in the
Minkowski bulk, and who is allowed to observe for an indefinite amount
of time all the way to the tip of the hat, at $i^+$ (see Fig. 3).  An
important feature of this boundary theory is that it includes a
Liouville field $L$ living in $\Sigma_2$.  This field accounts for the
time evolution in dual Minkowski bulk, and it has been argued that
conformal invariance is recovered in the limit of large $L$, which
corresponds to late times \cite{hat1,hat2}.  Another interesting
aspect of this picture is that the census taker would observe
collisions between the reference Minkowski bubble and other bubbles
that nucleate in its neighborhood. In this way, she would receive
information about other vacua, different from the ``parent" vacuum
where the Minkowski bubble nucleated. Hence, the field theory on
$\Sigma_2$ should encode a substantial amount of information about the
dynamics in the landscape of vacua.  It has been shown that,
accounting for bubble collisions, the surface $\Sigma_2$ can have any
genus, and so the boundary theory should include a sum over topologies
\cite{hat3}.

Based on the analogy with the black hole horizon complementarity \cite{complementarity},
it was proposed in \cite{hat1,hat2} that the dynamics of the entire multiverse may be
holographically dual to that of the causal patch of a single census taker.
The Euclidean theory on $\Sigma_2$ would then provide a complete description
of the multiverse.  We note however that there may be many different stable Minkowski vacua in
the landscape, and there seems to be no good reason to restrict
attention to the field theory associated with a particular one of
them. Also, it is not clear whether any such field theory on the
surface $\Sigma_2$ would fully represent the underlying dynamics,
since the census taker may not be able to see the full set of vacua.

In particular, there are some events in AdS whose future light cone is
completely engulfed by the big crunch, and those cannot be seen by a
Minkowski census taker.  Hence, it appears that we need to enlarge our
holographic screen in order to account for these regions. As mentioned
above, we propose that the bulk dynamics of AdS bubbles is encoded in
their boundary as well. Intuitively, this seems reasonable since we
know that the bulk of AdS spacetime does have a holographic
description.  Here, with AdS bubbles, the situation is somewhat
different because a future singularity develops.  It is conceivable
that this simply changes the boundary description
\cite{Hertog}, and we shall tentatively asume that this is
given in terms of degrees of freedom living in $\Sigma_2$ (as
mentioned above, for AdS bubbles $\Sigma_2$ is defined as the boundary
of the set of corresponding singular points at $c_+$).

\begin{figure}[ht]

\begin{center}\leavevmode  
\epsfxsize=16.5 cm
\epsfbox{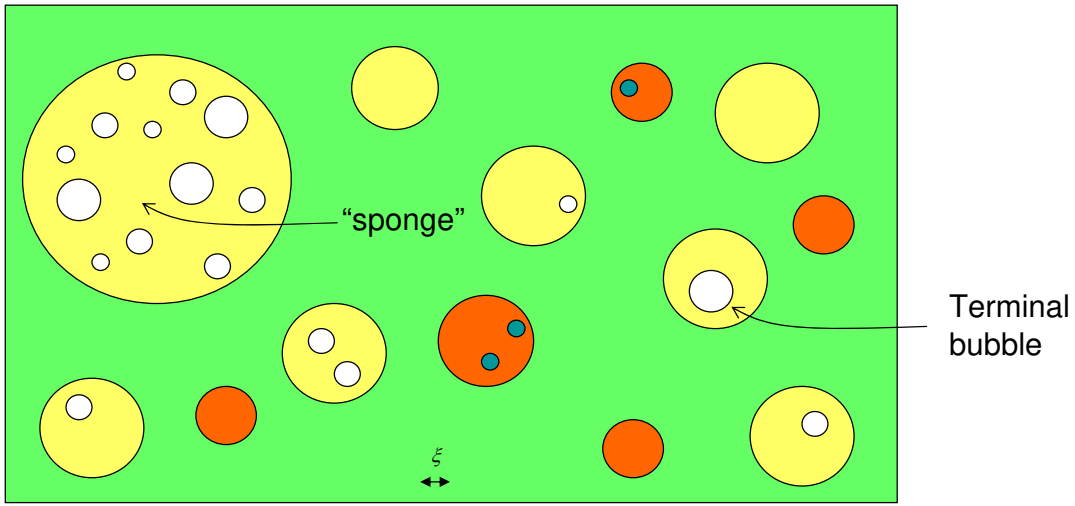}
\end{center}

\caption{For given co-moving resolution $\xi$, the eternal set $E$ can
  be mapped into the fiducial hypersurface $\Sigma_3$, and looks like
  a 3 dimensional space with holes in it. The holes correspond to
  geodesics in the congruence which have fallen inside of terminal
  bubbles.  The bulk dynamics of these bubbles is described in terms
  of degrees of freedom which live at the boundary of the holes. The
  images of bubbles of inflating vacua look like sponges, whose pores
  are occuppied by other inflating vacua, or by holes.}
\label{5j}
\end{figure}

\subsection{A conjecture}

Here, we propose that the dynamics of the landscape is encoded in a
Euclidean field theory defined on a set which includes all eternal
points in $c^+$, and not just those belonging to $\Sigma_2$.  The set
$E$ of eternal points is known to be a fractal \cite{aryal,fractal},
which we can represent on the fiducial hypersurface $\Sigma_3$.  For
any finite co-moving resolution $\xi$ this looks like a 3 dimensional
"sponge", with holes on it which correspond to the images of AdS and
Minkowski terminal bubbles, whose co-moving size is larger than
$\xi$ (see Fig. 4). Note that the points in the eternal set are mapped 
onto $\Sigma_3$, where they have a single image.  Hence, we have a
notion of co-moving distance in the eternal set given by the metric
$g_{ij}({\bf x})$ of the fiducial surface $\Sigma_3$.

Our conjecture is that the bulk dynamics of the eternally inflating
universe is represented by a Euclidean field theory living on that
sponge, where the resolution scale $\xi$ plays the role of a
renormalization scale.  The degrees of freedom of the field theory
live not only in the 3D "bulk" of the sponge, but also at the surfaces
$\Sigma_2$ which are at the boundary of the holes corresponding to the
images of all terminal bubbles. In this picture, we are invoking the
dS/CFT correspondence of Ref. \cite{strominger} for the inflating set,
while we are borrowing from the ideas of Refs. \cite{hat1,hat2,hat3} to
incorporate the bulk dynamics of terminal bubbles. For the reasons
explained in the previous Subsection, we are also including the
boundaries of AdS bubbles, in contrast with Refs. \cite{hat1,hat2,hat3}
where only the boundary of a single Minkowski bubble was considered.

Inside the "sponge" representing the inflating region, we will have
the images of bubbles of different inflating vacua, nested within each
other. Each one of these images will be thought of as an instanton in
the boundary theory \cite{hat1}.\footnote{The bubble walls of these
instantons may also carry some degrees of freedom, such as zero
modes, but these are unrelated to the holographic description of the
bubble interiors.}  Nested within the image $B_i$ of a bubble of
vacuum of type $i$, we will have the images of bubbles that nucleate
out of this vacuum later in time, and whose co-moving size is
therefore smaller. In this way, $B_i$ will itself look like a sponge,
whose ``pores" contain vacua of different types. Some of the ``pores"
may correspond to other inflating vacua, in which case they will
themselves look like sponges, and so forth. Of course, some of the
``pores" may correspond to Minkowski or AdS vacua, in which case they
are actual ``holes" in the eternal set (these would correspond to
bubbles of ``nothing").  As the
regulator is made finer and finer, we will see more and more of this
structure.

As mentioned above, in the eternal set we have a notion of co-moving
distance which is based on the metric $g_{ij}$ induced on an
arbitrarily chosen fiducial hypersurface $\Sigma_3$. A notion of
distance is in fact necessary in order to define the Wilsonian cut-off
$\xi$. If we change the surface, the metric changes, and we shall now
argue that this corresponds to Weyl rescalings in the
boundary theory.

Let us first discuss the case of de Sitter space.  We can consider a
pencil of geodesis of the standard congruence $C_0$ associated with
the flat coordinates of de Sitter (dS), in which 
\beq
ds^2=dt^2-e^{2Ht}d{\bf x}^2.  
\eeq
Suppose now we choose another congruence,
$C$.  Asymptotically, $C$ becomes comoving with $C_0$, so the surfaces
orthogonal to $C$ are nearly orthogonal to $C_0$ at late times.
Consider one such (late-time) surface $\Sigma_t$.  In terms of the
standard coordinates, this can be represented as 
\beq
\Sigma_t: \quad  t = f({\bf x}) ,
\eeq
where $f({\bf x})$ is a very slowly varying function.  The
asymptotic metric in the coordinates defined by the congruence $C$ is
\begin{equation}
ds^2 \approx dt^2 - e^{2Ht} e^{-2Hf({\bf x})} d{\bf
x}^2. \label{asympt}
\end{equation}
This shows that the change of congruence corresponds to a Weyl rescaling
of the metric at the future boundary.

In the multiverse, each bubble is a part of dS space, but, as we
described above, what remains of it asymptotically is just a
"sponge". A change of congruence will induce a Weyl
transformation in each sponge. In principle, from the argument above,
it is not clear whether the transformation is the same or not in
different sponges, because the Hubble rates $H$ are different in
different vacua.  However, this has to be the case because bubbles of
one of the vacua have to fit neatly into the "pores" of the
progenitor.\footnote{Suppose the parent and daughter vacua have
expansion rates $H$ and $H'$, respectively.  In flat de Sitter
coordinates they expand respectively as ${\rm exp}(Ht)$ and ${\rm
\exp}(H't')$.  If we chose the same origin for the time coordinate,
then the continuation of the surface $t = const$ of the parent vacuum
into a bubble is $t' = (H/H')t$.  This matching of the exterior and
interior regions gives a surface of constant scale factor; it is
easy to see that the corresponding rescaling is indeed
continuous across the bubble boundaries.}

Weyl rescalings can be used in order to change the size of
instantons. For a given congruence, by choosing the fiducial surface
to be at a later time, the size of all instantons becomes
bigger. However, because of the self-similar structure of the bulk
attractor at late times, the result of this rescaling
will not change the distribution of sizes and of bubbles of different
types at the boundary.  Thus, self-similarity at late times
corresponds to UV scale invariance in the boundary theory.
On the other hand, if we look at the boundary theory on the largest possible
scales, the distribution of instantons of different vacua will reflect
the initial conditions of inflation, and will not have this
invariance.\footnote{There is some evidence that the theory is invariant 
not just under rescalings, but under the full conformal group. This will be discussed 
elsewhere \cite{conformal}.}

\section{The boundary measure}

Here, we propose that the natural way to address the measure problem
of inflationary cosmology is to use the Wilsonian cutoff of the
boundary theory.

Formally, we may consider the amplitude,
\begin{equation}
Z[\bar\phi({\bf x})]=\int D\phi\ e^{iS[\phi]}. \label{partition}
\end{equation}
Here $S$ is the bulk action, and the integral is over bulk fields
$\phi$ approaching the prescribed $\phi={\bar\phi}({\bf x})$ at the
future boundary. Bulk fields do not really approach constant values at late
times, since in particular their values keep changing due to bubble nucleation (as well
as due quantum 
fluctuations of light fields). In order to make sense of Eq. (\ref{partition}), bulk fields should be smeared over 
a fixed co-moving  scale $\xi$, and likewise for the boundary values $\bar\phi$. 
With this coarse-graining,
the values of the fields are frozen after the co-moving wavelength $\xi$ crosses the horizon,
and the boundary condition can be implemented.
We may use the amplitude (\ref{partition}) in order to compute
correlators. For instance, the two point function is given by
\begin{equation}
\langle \bar\phi({\bf x})\ \bar\phi({\bf x'})\rangle =\int D\bar\phi\ 
\bar\phi({\bf x})\ \bar\phi({\bf x'})
\left|Z[\bar\phi]\right|^2.
\end{equation}
In principle, we should also specify boundary conditions for $\phi$ on
some initial fiducial surface, such as the $\Sigma_3$ we have
discussed above. However, based on the arguments we presented in the
previous Section, we expect that the initial boundary condition will
only determine the infrared behaviour of correlators. On the other
hand, we are interested in the UV fixed point (corresponding to the
attractor behaviour in the bulk description), and the initial boundary
condition will not play a role for our present purposes.  Hence, we
shall simply omit it in the following discussion. By analogy with
AdS/CFT, we now posit that the bulk dynamics is dual to a Euclidean
theory living at the boundary, where now the $\bar\phi({\bf x})$ play
the role of sources for operators in the boundary theory. The
conjecture is that (\ref{partition}) is also given by
\begin{equation}
Z[\bar\phi({\bf x})]=e^{i W_{CFT}[\bar\phi]},
\end{equation}
where $W_{CFT}$ is the effective action for a boundary field theory with
appropriate couplings to the external sources $\bar\phi$. If the theory is
regularized with a cutoff $\xi$, then we should think of the
configurations $\bar\phi({\bf x})$ as coarse-grained on the scale
$\xi$.

We propose that in order to determine the probabilities of given
semiclassical processes in the bulk, we should do the counting in the
regularized boundary theory, where this counting will be finite. The
idea is that any bulk process will also be represented in the boundary
theory. In the coarse grained description, only a finite number of
these processes will be resolved, and relative probabilities can
be defined as the ratios of occurrences in the limit $\xi\to 0$.

Let us now argue that this definition of the measure is closely
related to the so called scale factor cutoff measure, where we take
into account only those processes which happen before a fixed scale
factor time $a_c$, and then we take the limit where $a_c \to \infty$
in order to determine the probabilities for the processes to occur.
Suppose we are interested in 4D bulk processes which require a
resolution corresponding to the physical length scale
$\lambda_{min}$. For example, if we have in mind some cosmological
process at a given scale, we may think of $\lambda_{min}$ as a
somewhat smaller scale, just so that the process can be properly
identified.  Now, the co-moving wavelength
$\xi a$ corresponding to the boundary cut-off will be smaller than $\lambda_{min}$ 
provided that the process
takes place at sufficiently early times,
\begin{equation}
a < a_c = \lambda_{min}/\xi.       \label{iruv}
\end{equation}
This equation relates the Wilsonian cutoff $\xi$ of the
boundary theory to the scale factor cutoff $a_c$.  Note that the
infrared limit in the bulk theory, $a_c\to \infty$, corresponds to the
UV limit in the boundary theory, $\xi\to 0$.\footnote{Note that the
resolution scale $\lambda_{min}$ can be thought of as a Wilsonian 
UV cutoff in the bulk theory.} This relation is of course familiar from the
analogous context of AdS/CFT. There, the RG flow is associated with
radial displacement in the bulk, whereas here it is associated with
scale-factor time evolution.
\footnote{As noted in the text, a connection between time evolution and RG
flow is to be expected by analogy with AdS/CFT, and had already been observed
e.g. in Refs. \cite{strominger,hat1,hat2}. Here, we are making the connection more
precise, by relating the RG flow to scale factor time evolution (as
opposed to, say, proper time evolution).}

The argument is somewhat more involved for terminal bubbles.  The
interior of these bubbles looks like an open FRW universe, 
\beq
ds^2 = d\tau^2 - a_{FRW}^2(\tau)(d\zeta^2 + \sinh^2\zeta d\Omega^2),
\eeq
and the corresponding boundary theory lives at space-like infinity of
the 3-dimensional space-like hyperboloids that foliate this universe,
$\zeta\to\infty$.  The size of the images on the holographic screen is
determined by the value of the scale factor $a_{FRW}$, as well as the
radial distance $\zeta$ to the center of the hyperboloid.  For a
terminal bubble nucleated in parent vacuum $i$ at a scale factor
$a_{nuc}$, the comoving radius of its future boundary is $R=(H_i
a_{nuc})^{-1}$.  The regulator scale $\xi$ applied to this boundary
subtends an angle $\theta_\xi \approx \xi/R$ from the bubble center,
and the corresponding physical distance on a hypersurface of constant
$\tau$ at radial coordinate $\zeta$ is
\beq
d_\xi(\zeta,\tau) = \theta_\xi a_{FRW}(\tau)\sinh\zeta.
\eeq
(We assume that $\xi\ll R$.)  Requiring that $d_\xi$ is smaller than
the resolution scale $\lambda_{min}$, we have
\beq
H_i a_{nuc}a_{FRW}(\tau)\sinh\zeta < \lambda_{min}/\xi.
\label{1}
\eeq
Now, for $\zeta \gg 1$, the expression on the left-hand side is
precisely the scale factor $a$ in the bubble interior
\cite{VW,youngness2,Bousso:2008hz}.  The factors
$a_{nuc}$, $\sinh\zeta$ and $H_i a_{FRW}(\tau)$ account respectively
for the expansion from the fiducial hypersurface to nucleation, from
nucleation to the time when the geodesic at a given $\zeta$ crosses
the bubble wall, and for the expansion inside the bubble.  (Note that
with the definitions we adopted, $a_{nuc}$ is dimensionless, while
$a_{FRW}$ has the dimension of length.)  Thus, we recover
Eq.~(\ref{iruv}), that is, the scale factor cutoff.\footnote{
Throughout the paper, we disregard focusing of geodesics at domain
wall crossing and the ambiguities associated with continuing geodesic
congruences into the ``fuzzy'' quantum regions in the vicinity of
bubble nucleation events.  These issues require further study.}

The correspondence between the boundary and scale factor
measures is nonetheless only approximate, and it can break down when we are
interested in processes involving wavelengths much smaller than the
Hubble radius (as is often the case). The physical reason is simple:
while their physical size is smaller than the Hubble radius,  these
modes can be affected by all sorts of other subhorizon processes and
need not simply evolve by conformal stretching with the expansion of
the universe.\footnote{A similar situation is encountered in AdS/CFT, where
wavelengths which are smaller than the AdS curvature radius are {\em not} 
trivially mapped into the boundary theory.}

A related observation is that all finite co-moving wavelenths at the
future boundary correspond to frozen modes (since the co-moving size
of the horizon shrinks to zero asymptotically).  This means that the
coarse grained configurations $\bar\phi(\bf{x})$, from which we must
retreive the information about bulk events, are configurations which
are frozen in with the expansion.  Physically, it is not surprising
that the information gets to future infinity in the form of long
wavelength modes. The events we are trying to reconstruct will give
away, say, electromagnetic or gravitational radiation whose wavelength
is conformally stretched. More generally, information is bound to leak
from short to long wavelengths through interactions.  Once
the information is in the form of wavelengths bigger than the horizon,
it becomes indestructible: no causal process can erase it. 
However, the precise way in which the information {about bulk events} 
is encoded at infinity can be more complicated than just conformal stretching from
small scales, and so the correspondence with scale factor cutoff is
not exact.

In fact, the scale factor measure itself is not uniquely defined on
sub-horizon scales.  In regions of structure formation, where
geodesics converge and cross, the scale factor is not a good time
variable, and this leads to ambiguities \cite{DGSV,Bousso:2008hz}.
The scale factor may start decreasing along some geodesics,
until it vanishes at a caustic, and then start increasing again.  A
given value of the scale factor may thus be reached multiple times as
we move along a geodesic, and it is not clear which of these
occurrences should be used to implement the cutoff. In Ref.\cite{DGSV}
it was proposed that the first occurrence should be used. On the other
hand, it was pointed out in \cite{Bousso:2008hz} that the resulting
cutoff surface is strongly influenced by the local details of
structure formation.  It has a rather ``spiky'' appearance and is not
generally spacelike.  An alternative possibility, indicated in
\cite{Bousso:2008hz} is to define the geodesic congruence using the
spacetime metric smoothed over some characteristic scale.  If this
scale is chosen to be larger than the typical scale of structure
formation, then the congruence will always be diverging, except in the
collapsing AdS regions.  In AdS regions, the scale factor will reach
some maximum value $a_{max}$ and then decrease down to zero, so
additional prescriptions are needed to handle this case. If the cutoff
value is $a_c < a_{max}$ this value will occur twice on the geodesic,
and hence it is not very clear how to implement the scale factor
cutoff.

Information about an event in the 4-dimensional spacetime travels to
the future infinity along null and timelike geodesics.  In the case of
events occurring in a dS vacuum, such as perhaps our own, this
information is represented in a region within the comoving horizon of
that event on the future boundary. For example, one can expect that
the collapse of a protocloud resulting in galaxy formation will be
encoded in the field values of the entire comoving horizon region,
rather than being localized near the comoving location of the
galaxy. This seems to suggest that the boundary measure is not likely
to be influenced by the local situation in the vicinity of the galaxy,
as it would be with the version of the scale factor cutoff measure
adopted in \cite{DGSV}.  Instead, we may expect that in this case the
boundary measure will be well approximated by the scale factor cutoff,
with a geodesic congruence based on the metric smoothed on the scale
of the horizon.

\section{Conclusions and discussion}

We have argued that the dynamics of the inflationary multiverse may
have a dual description in the form of a lower dimensional Euclidean
field theory defined at the future infinity.  The measure of the
multiverse can then be defined by imposing a Wilsonian ultraviolet
cutoff $\xi$ in that theory.  In the limit of $\xi\to 0$, the boundary
theory becomes conformally invariant, approaching a UV fixed point.

On super-horizon scales, the UV cutoff $\xi$ corresponds to an
infrared (late time) scale factor cutoff in the bulk theory, and the
renormalization group flow corresponds to the scale factor time
evolution.  The asymptotic scale invariance of the boundary theory is
reflected in the late-time attractor behavior of eternal inflation.
The correspondence between the boundary measure and scale factor
cutoff is not precise on sub-horizon scales, but it is expected to
hold approximately if the scale factor is defined using the metric
suitably averaged over the horizon.

The proposal we have outlined in this paper is only a sketch of the
boundary measure, with a number of open questions left for future
research.  One of these questions is related to the geodesic crossing.
In order to avoid geodesic crossing, we assumed that our geodesic
congruences are constructed from a metric averaged over the structure
formation scale. We have also ignored the focusing of geodesics as
they go through domain walls separating different phases. 
Such approximations appear to be out of place in a fundamental theory.   
In fact, it is not clear to what extent geodesic congruences are
necessary for our construction.  They are of course necessary to
establish the correspondence with the scale factor measure, but one
can hope that the duality between the bulk and boundary theories and
the boundary measure can be formulated entirely in the framework of
field theory, without reference to geodesic congruences.

The future infinity, where the boundary theory is defined, is a
fractal set consisting of infinitely fine ``sponges'' representing
different inflating vacua.  At any finite resolution $\xi$, each
sponge is a $3D$ manifold, which is bounded by its borders with other
sponges and by the boundaries of terminal bubbles.  The latter
boundaries, as well as the sponges themselves, are sites of
(asymptotically) conformal field theories; the corresponding cental
charges have been estimated in \cite{strominger,hat2}.  In the limit of $\xi\to
0$, the sponges become self-similar fractals of dimension $d_S < 3$
and their boundaries become fractals of $d_B > 2$, due to bubble
collisions.  (For low bubble nucleation rates, which is usually the
case, $d_S$ and $d_B$ are very close to 3 and 2, respectively.)  It
would be very interesting to see if the boundary theory can be defined
directly on this fractal set.  If so, it will have to be a rather
unconventional field theory.

We note, finally, that the detailed dynamics of the boundary theory
may not be needed in order to apply the boundary measure. The
advantage of duality is precisely that calculations can be done in the
bulk, where the theory is weakly coupled. To make use of this
procedure, it suffices to find out what are the asymptotic co-moving
wavelenghts carrying the information about the process of our
interest. Technically, this may be more or less complicated depending
on the process. But in principle, one should be able to determine it
from standard bulk physics.

\section{Acknowledgements}

We would like to thank Leonard Susskind and Tomeu Fiol for useful discussions.  
We are also grateful to Misao Sasaki and the Yukawa Institute for their
hospitality at the workshop YITP-W-07-10, where this work was
originated.  This work was supported in part by the Fundamental
Questions Institute (JG and AV), by grants FPA2007-66665C02-02 and
DURSI 2005-SGR-00082 (JG), and by the National Science Foundation
(AV).

\end{document}